\documentclass[pra,twocolumn]{revtex4}%
\usepackage{amsfonts}
\usepackage{amsmath}
\usepackage{amssymb}
\usepackage{graphicx}%
\providecommand{\U}[1]{\protect\rule{.1in}{.1in}}

\begin{document}
\title{Heralded quantum memory for single-photon polarization qubits }
\author{G. W. Lin$^{1}$}
\author{X. B. Zou$^{1}$}
\email{xbz@ustc.edu.cn}
\author{X. M. Lin$^{2}$}
\email{xmlin@fjnu.edu.cn}
\author{G. C. Guo$^{1}$}
\affiliation{$^{1}$Key Laboratory of Quantum Information, Department of Physics, University
of Science and Technology of China, Hefei 230026, People's Republic of China}
\affiliation{$^{2}$School of Physics and Optoelectronics Technology, Fujian Normal
University, Fuzhou 350007, People's Republic of China}


\begin{abstract}
We propose a scheme to implement a heralded quantum memory for single-photon
polarization qubits with a single atom trapped in an optical cavity. In this
scheme, an injected photon only exchanges quantum state with the atom, so that
the heralded storage can be achieved by detecting the output photon. We also
demonstrate that the scheme can be used for realizing the heralded quantum
state transfer, exchange and entanglement distribution between distant nodes.
The ability to detect whether the operation has succeeded or not is crucial
for practical application.

\end{abstract}
\maketitle

\emph{Introduction}.---A quantum memory, a storage device that can faithfully
store and retrieve quantum state of traveling light pulse, is requisite to
long-distance quantum communication networks and distributed quantum computers
\cite{Kimble,Briegel,Duan,Cirac,D. Boozer,Zhao,Jiang,Kok,Brassard}. By
following the seminal protocol proposed by Duan et al. \cite{Duan},
significant advance has been made in storing single photon by using atomic
ensembles as storage mediums
\cite{Chou,Matsukevich,Chane,Eisaman,Black,Thomp,Choi,Karpa,B.
Julsg,Julsgaa,Takano,Alexey,Chuu,Honda,Appel,Chen}. Early experiments using
atomic ensembles mainly focused on the storage and retrieval of single photon
with the fixed polarization \cite{Chane,Eisaman}. Recent experiments show that
single-photon polarization qubits can be stored in two atomic ensembles
\cite{Choi,Karpa}. However, above studies are thwarted by photon losses since
one can't determine whether or not the incoming photon has been stored or
lost. Fortunately, such unpredictable failure may be largely remedied by a
heralding feature that announces photon arrival and successful storage without
destroying the stored quantum state. In Ref. \cite{Haruka}, a heralded storage
has been achieved by means of a spontaneous Raman process that can absorb a
single-photon with arbitrary polarization and simultaneously emit a photon
with fixed polarization. However, due to small spontaneous Raman scattering
probability, the heralded storage occurs rarely, while its polarization state
is restored with high fidelity. Cavity quantum electrodynamics (QED) provides
another ideal interface between atoms and photons \cite{Wilk,Wang,Cirac,van
Enk,Boozer,Fleischhauer,Lloyd}. In the initial proposal for the implementation
of quantum networks in cavity QED \cite{Cirac}, the quantum information of
photons is encoded in the Fock basis, i.e., the zero- and one-photon Fock
states \cite{Cirac,Boozer}. Ref. \cite{Lloyd} proposed a robust method for
transmitting entangled polarization state over long distances and
teleportation of atomic state via measurements of all four Bell states, using
a novel method of sequential elimination.

In this paper, we propose a heralded quantum memory for arbitrary polarization
state with a single atom trapped in an optical cavity. Our scheme is based on
quantum-state swap between single-photon pulse and trapped atom. The heralded
storage can be achieved by detecting the output photon. Numerical simulation
results show that our scheme has a high success probability and the retrieved
photon has so well-defined waveforms that it is easy to interfere with other
photons. We also show that the scheme can allow the heralded quantum state
transfer, exchange and entanglement distribution between distant nodes. The
ability to detect whether the operation has succeeded or not, is crucial for
practical application \cite{Briegel,Haruka}.

\emph{The building block and numerical simulations}.---As shown in Fig.1(a),
our model consists of a single atom inside a one-sided optical cavity.
\begin{figure}[ptb]
\includegraphics[width=2.5in]{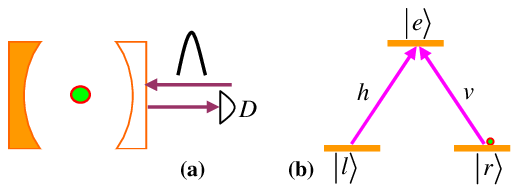}\newline\caption{(Color online) (a)
Schematic setup to reflect a single photon pulse from one-sided cavity. (b)
The relevant atomic level structure and transitions.}%
\label{1}%
\end{figure}The relevant atomic levels are depicted in Fig. 1(b), the states
$\left\vert l\right\rangle $ and $\left\vert r\right\rangle $ correspond to
two stable ground states, and $\left\vert e\right\rangle $ denotes an excited
state. The cavity supports two degenerate cavity modes $a^{h}$ and $a^{v}$
with different polarizations h and v. We assume that the transitions
$\left\vert l\right\rangle \leftrightarrow\left\vert e\right\rangle $ and
$\left\vert r\right\rangle \leftrightarrow\left\vert e\right\rangle $ are
resonantly coupled to two cavity modes $a^{h}$ and $a^{v}$ with the strength
$g_{h}$ and $g_{v}$, respectively. The interaction between the atom and cavity
modes is described by $H_{I}=g_{h}\left\vert e\right\rangle \left\langle
l\right\vert a^{h}+g_{v}\left\vert e\right\rangle \left\langle r\right\vert
a^{v}+H.c.$. The time evolutions of cavity modes $a^{h}$ and $a^{v}$ are given
by \cite{Cirac,Gardiner,Duan3}
\begin{align}
\dot{a}^{h(v)}\left(  t\right)   &  =-i\left[  a^{h(v)}\left(  t\right)
,H_{I}^{^{\prime}}\right]  -\frac{\kappa}{2}a^{h(v)}\left(  t\right)
\nonumber\\
&  -\sqrt{\kappa}a_{in}^{h(v)}\left(  t\right)  ,
\end{align}
where $H_{I}^{^{\prime}}=H_{I}-i\frac{\gamma_{e}}{2}\left\vert e\right\rangle
\left\langle e\right\vert $ \cite{Fleischhauer,Duan4}, $\kappa$ is the cavity
decay rate for two cavity modes, and $\gamma_{e}$ is spontaneous-emission rate
of the excited state. The output operator $a_{out}^{h(v)}\left(  t\right)  $
is connected with the input operator $a_{in}^{h(v)}\left(  t\right)  $ by the
input-output relation $a_{out}^{h(v)}\left(  t\right)  =a_{in}^{h(v)}\left(
t\right)  +\sqrt{\kappa}a^{h(v)}\left(  t\right)  $.

In this paper, two cases are considered: (i) the atom is prepared in the state
$\left\vert l\right\rangle $ ( or $\left\vert r\right\rangle $), and a
v-polarized (h-polarized) photon $\left\vert v\right\rangle $ ($\left\vert
h\right\rangle $) is injected. In this case, the Hamiltonian $H_{I}^{^{\prime
}}$ does not work and the injected photon sees an empty cavity. Thus, the
polarization of the input photon is not changed and we have
\cite{Gardiner,Duan3}
\begin{equation}
a_{out}\left(  \omega\right)  =\frac{-\kappa/2-i\delta}{\kappa/2-i\delta
}a_{in}\left(  \omega\right)  ,
\end{equation}
where we assume that the center frequency of input photon pulse $\omega_{0}$
is resonant with the cavity mode and $\delta=\omega-\omega_{0}$. If $\kappa
\gg\delta$, we have $a_{out}\left(  \omega\right)  \approx-a_{in}\left(
\omega\right)  $. Next we consider the case (ii): the atom is prepared in the
state $\left\vert l\right\rangle $ ( or $\left\vert r\right\rangle $), and a
h-polarized (v-polarized) photon $\left\vert h\right\rangle $ ($\left\vert
v\right\rangle $) is injected. In this case, taking the adiabatic limit
\cite{Fleischhauer,Duan4}, we can obtain the input-output relation
\begin{equation}
a_{out}\left(  \omega\right)  =a_{out}^{h(v)}\left(  \omega\right)
+a_{out}^{v(h)}\left(  \omega\right)  ,
\end{equation}
with
\begin{equation}
a_{out}^{h(v)}\left(  \omega\right)  =[1-\frac{\kappa}{\kappa/2-i\delta
+2g_{h(v)}^{2}\kappa/(4g_{v(h)}^{2}+i\kappa\gamma_{e})}]a_{in}^{h(v)}\left(
\omega\right)  ,
\end{equation}
and
\begin{equation}
a_{out}^{v(h)}\left(  \omega\right)  =\frac{g_{h(v)}\kappa/g_{v(h)}}%
{[\kappa/2-i\delta+g_{h(v)}^{2}\kappa/(2g_{v(h)}^{2}+i\kappa\gamma_{e}%
/2)]}a_{in}^{h(v)}\left(  \omega\right)  .
\end{equation}
It is seen from Eq. (3-5) that if conditions $\kappa\gg\delta$, $4g_{v(h)}%
^{2}\gg\kappa\gamma_{e}$, and $g_{h}\approx g_{v}$ are satisfied, we have
$a_{out}^{v(h)}\left(  \omega\right)  \approx$ $a_{in}^{h(v)}\left(
\omega\right)  $. Therefore, the polarization of the input photon is changed
and we realize state flip operation $\left\vert l\right\rangle \left\vert
h\right\rangle \leftrightarrow\left\vert r\right\rangle \left\vert
v\right\rangle $.

Now, we give a detailed analysis of this building block through numerical
simulation with the Hamiltonian \cite{Fleischhauer,Duan4}
\begin{align}
H  &  =H_{I}^{^{\prime}}+%
{\displaystyle\sum\limits_{\digamma=h,v}}
{\displaystyle\int\nolimits_{-\omega_{b}}^{+\omega_{b}}}
\omega d\omega\Theta_{\digamma}^{\dagger}(\omega)\Theta_{\digamma}%
(\omega)\nonumber\\
&  +i\sqrt{\frac{\kappa}{2\pi}}%
{\displaystyle\sum\limits_{\digamma=h,v}}
{\displaystyle\int\nolimits_{-\omega_{b}}^{+\omega_{b}}}
d\omega\lbrack a^{\digamma}\Theta_{\digamma}^{\dagger}(\omega)-a^{\digamma
\dagger}\Theta_{\digamma}(\omega)],
\end{align}
where $\Theta_{\digamma}(\omega)$ with the standard relation $[\Theta
_{\digamma}(\omega),\Theta_{\digamma}^{\dagger}(\omega^{^{\prime}}%
)]=\delta(\omega-\omega^{^{\prime}})$ denotes the one-dimensional free-space
mode coupled to the cavity modes.

Assuming that the input single-photon pulse is Gaussian pulse of the form
$f\left(  t\right)  \propto\exp\left[  -\left(  t-\frac{T}{2}\right)
^{2}/\left(  \frac{T}{5}\right)  ^{2}\right]  $, where $T$ is the pulse
duration. We first consider above-mentioned case (ii) that a polarized photon
$\left\vert h\right\rangle $ ($\left\vert v\right\rangle $) is injected into
cavity and the atom is prepared in state $\left\vert l\right\rangle $
$(\left\vert r\right\rangle )$. Let $P$ be the success probability for the
flip $\left\vert l\right\rangle \left\vert h\right\rangle \leftrightarrow
\left\vert r\right\rangle \left\vert v\right\rangle $. In Fig. 2, we plot $P$
versus the normalized cavity coupling rate $g/\kappa$, assuming $\gamma
_{e}=\kappa$, with different the pulse width: $\kappa
T=\{10,20,30,40,50,60,120\}$. From Fig. 2, \begin{figure}[ptb]
\includegraphics[width=3.0in]{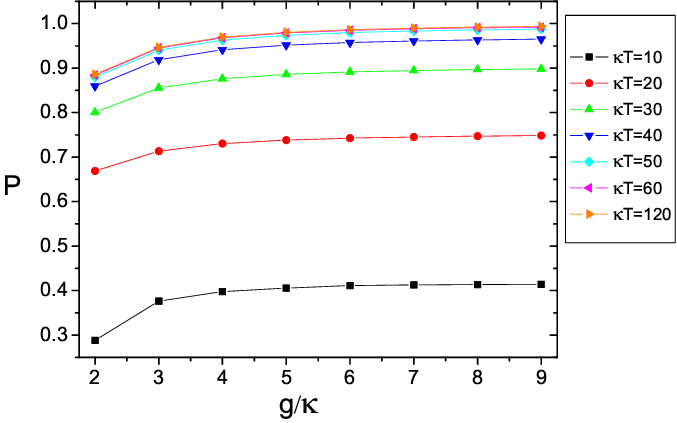}\newline\caption{(Color online) The
success probability P for the flip $\left\vert l\right\rangle \left\vert
h\right\rangle \leftrightarrow\left\vert r\right\rangle \left\vert
v\right\rangle $ as a function of $g/\kappa$ for $\kappa
T=\{10,20,30,40,50,60,120\}$, on the assumption that $\gamma_{e}=\kappa$ and
$g_{h}\bigskip=g_{v}$=g.}%
\label{a}%
\end{figure}we see that when $g\geq2\kappa$ and $\kappa T$ $\geq60$ the
success probability can be up to $90\%.$ \begin{figure}[ptb]
\includegraphics[width=3.0in]{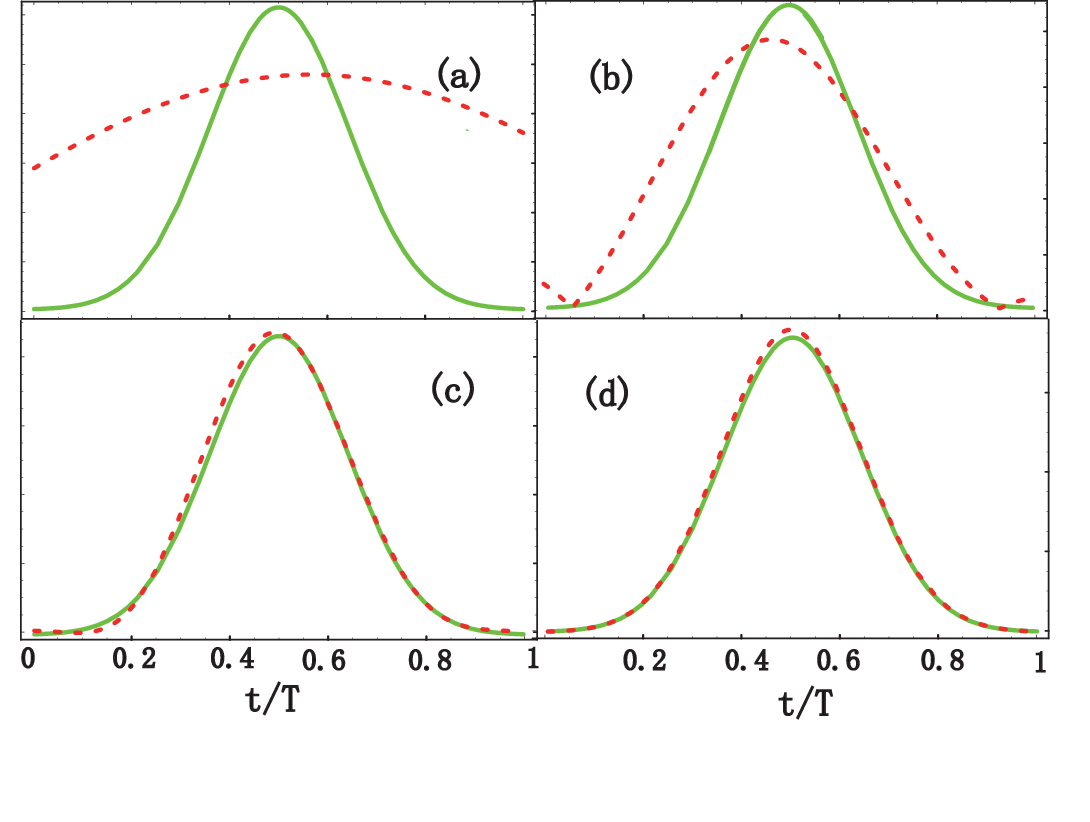}\newline\caption{(Color online) The
shape functions $|f(t)|$ for the input pulse (solid curve) and the output
pulse (dashed curve) for the case $\kappa T$ = (a) 10, (b) 30, (c) 60, and (d)
90. Other common parameters: $\gamma_{e}=\kappa$, $g_{h}\bigskip=g_{v}$=g, and
$g=2\kappa$.}%
\label{aa}%
\end{figure}Fig. 3 shows that when $\kappa T$ $\geq60$ the output pulse shapes
$|f(t)|$ almost completely overlap with the input pulse shapes. Then, we
consider the case (i) that a polarized photon $\left\vert h\right\rangle $
($\left\vert v\right\rangle $) is injected into cavity and the atom is in the
state $\left\vert r\right\rangle $ $(\left\vert l\right\rangle )$. In this
case, the atom is decoupled from the cavity field, thus the influence of the
atomic spontaneous emission can be ignored and the dominant noise is the
distortion between the output and the input pulse. In Fig. 4, we plot the
output and the input pulses shapes with different the pulse width: $\kappa
T=\{30,60,90,120\}$.

\begin{figure}[ptb]
\includegraphics[width=3.0in]{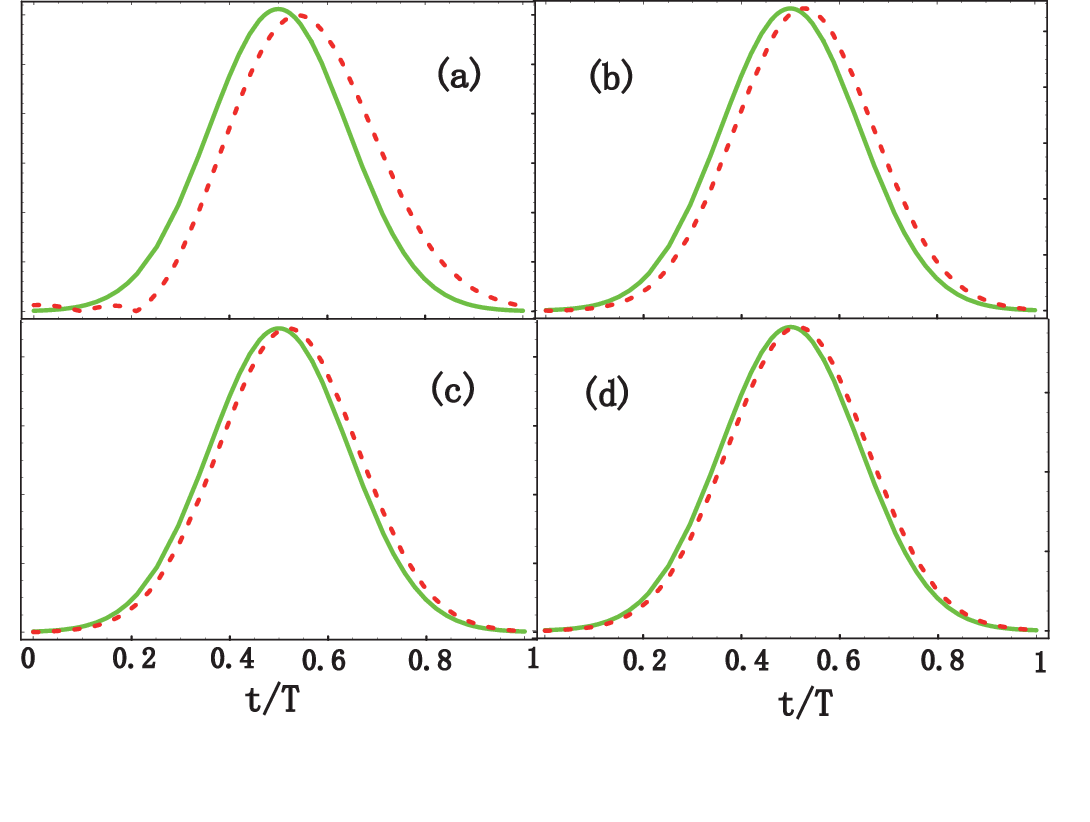}\newline\caption{(Color online) The
shape functions $|f(t)|$ for the input pulse (solid curve) and the output
pulse (dashed curve) for the case $\kappa T$ = (a) 30, (b) 60, (c) 90, and (d)
120.}%
\label{s}%
\end{figure}\emph{Quantum swap gate and Quantum memory}.---Assume that the
injected photon and the atom are initially in arbitrary superposition of two
polarized states and two ground states, respectively, i.e., $(\alpha\left\vert
h\right\rangle +\beta\left\vert v\right\rangle )\otimes(\zeta\left\vert
l\right\rangle +\eta\left\vert r\right\rangle ),$ where coefficients $\alpha$,
$\beta$, $\zeta$, and $\eta$ satisfy relations $\left\vert \alpha\right\vert
^{2}+\left\vert \beta\right\vert ^{2}=1$ and $\left\vert \zeta\right\vert
^{2}+\left\vert \eta\right\vert ^{2}=1$. After the photon pulse is reflected
by the cavity, the state of total system in the ideal case will evolve into
$(-\eta\left\vert h\right\rangle +\zeta\left\vert v\right\rangle
)\otimes(-\beta\left\vert l\right\rangle +\alpha\left\vert r\right\rangle )$,
which corresponds to quantum state exchange operation between photon and atom,
apart from phase factors that can be eliminated by the appropriate subsequent
logic operations \cite{Lin}. We quantify the quality of swap gate through a
numerical simulation, and the parameters are referred to Ref. \cite{Sauer},
i.e., $(g_{0},\kappa,\gamma_{e})/2\pi=(27,4.8,6)MHz$. Suppose that the initial
state of system is given by $\left\vert \Psi(0)\right\rangle =(\left\vert
h\right\rangle +\left\vert v\right\rangle )\otimes(\left\vert l\right\rangle
+\left\vert r\right\rangle )/2$, Fig. 5 shows that the swap gate has a high
fidelity and the variation of fidelity is about 0.01 for g varying from
$g_{0}$ to $g_{0}/2$.

\begin{figure}[ptb]
\includegraphics[width=3.0in]{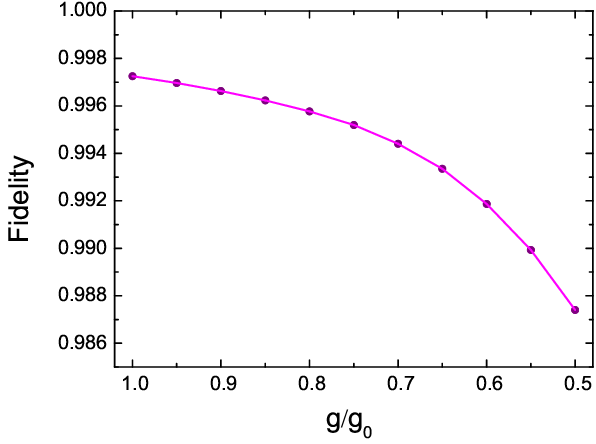}\newline\caption{(Color online) The
fidelity of the quantum state exchange operation with $g/g_{0}$. Here we have
assumed a Gaussian shape for the input pulse with $f\left(  t\right)
\propto\exp[ -(t-\frac{T}{2})^{2}/(\frac{T}{5})^{2}] $, and chosen the
parameters $T=5\mu s$, $(g_{0},\kappa,\gamma_{e})/2\pi=(27,4.8,6)MHz$,
$g_{h}\bigskip=g_{v}$=g.}%
\label{2}%
\end{figure}

Based on above-mentioned swap operation, we consider the heralded quantum
storage. Suppose that a photon is initially in arbitrary polarized state
$(\alpha\left\vert h\right\rangle +\beta\left\vert v\right\rangle )$ and atom
is prepared in the state $\left\vert r\right\rangle $. After the photon pulse
is reflected by the cavity, the state of system becomes
\begin{equation}
(\alpha\left\vert h\right\rangle +\beta\left\vert v\right\rangle
)\otimes\left\vert r\right\rangle \rightarrow\left\vert h\right\rangle
\otimes(\beta\left\vert l\right\rangle -\alpha\left\vert r\right\rangle ).
\end{equation}
Detection of the reflected photon in the state $\left\vert h\right\rangle $
heralds the mapping of the input polarization state onto atom. We note that
Fleischhauer et al. have proposed an interesting scheme for storing
single-photon quantum states by adiabatic evolution of dark states, which has
a high success probability, but not be heralded \cite{Fleischhauer}. For
transmitting entangled polarization state over long distances in realistic
noisy channels, Ref. \cite{Lloyd} proposed a robust method for capturing
photon in optical cavities, and storing it in atoms, in which they detected if
the atom has jumped out of the initial state, i.e., absorbed a photon, using
the fluorescence technique by driving a cycling transition from initial state
of the atom to an accessorial excited state.

The approach for retrieving photonic states is similar to that for storing
process. A photon pulse in the state $\left\vert h\right\rangle $ is reflected
by the cavity, the state evolution of system can be represented by
\begin{equation}
\left\vert h\right\rangle \otimes(\beta\left\vert l\right\rangle
-\alpha\left\vert r\right\rangle )\rightarrow(\alpha\left\vert h\right\rangle
+\beta\left\vert v\right\rangle )\otimes\left\vert r\right\rangle .
\end{equation}
After exchanging quantum states, the photonic qubit has been successfully
retrieved and leaving atom in initial state $\left\vert r\right\rangle $ for
storage of another photon.

\emph{Quantum communication protocols}.---One application of our scheme is the
heralded quantum state transfer and exchange between two atomic memories are
shown in Fig. 6(a). \begin{figure}[ptb]
\includegraphics[width=3.3in]{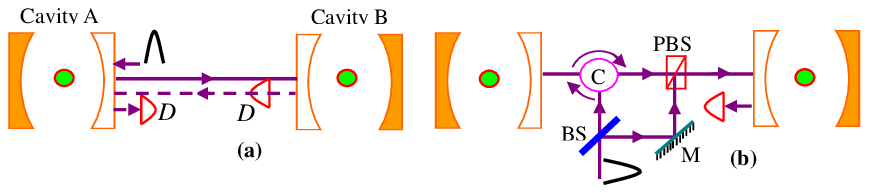}\newline\caption{(Color online) (a)
Quantum state transfer from cavity A to cavity B. (b) Creation of entanglement
between two atomic memories. Where D, PBS, C, BS, and M denote single photon
detector, polarizing beam splitter, circulator, 50/50 beam splitter, and
mirror, respectively.}%
\label{6}%
\end{figure}Suppose that the initial state of a single photon pulse and two
atomic memories is $\left\vert \Phi(0)\right\rangle =\left\vert h\right\rangle
\otimes(\alpha\left\vert l\right\rangle _{A}+\beta\left\vert r\right\rangle
_{A})\otimes\left\vert l\right\rangle _{B}$, where $\left\vert \theta
\right\rangle _{\chi}(\theta=l,r$, $\chi=A,B)$ denotes quantum state of the
atom trapped in cavity $\chi$. The photon pulse is sequentially reflected by
the cavities A and B, the state of the combined system in the ideal case will
evolve into $\left\vert v\right\rangle \otimes\left\vert r\right\rangle
_{A}\otimes(\alpha\left\vert l\right\rangle _{B}+\beta\left\vert
r\right\rangle _{B})$. Detection of the photon in the state $\left\vert
v\right\rangle $ heralds that quantum state transfer from cavity A to cavity B
has been performed successfully. If the initial state of the photon-atom
system is prepared in $\left\vert \Phi^{^{\prime}}(0)\right\rangle =\left\vert
h\right\rangle \otimes(\alpha\left\vert l\right\rangle _{A}+\beta\left\vert
r\right\rangle _{A})\otimes(\zeta\left\vert l\right\rangle _{B}+\eta\left\vert
r\right\rangle _{B})$, the photon pulse is sequentially reflected by the
cavities A and B, then reflected by the cavity A again. the state of combined
system will evolve into $\left\vert h\right\rangle \otimes(\zeta\left\vert
l\right\rangle _{A}+\eta\left\vert r\right\rangle _{A})\otimes(\alpha
\left\vert l\right\rangle _{B}+\beta\left\vert r\right\rangle _{B})$. After
the detection of the photon in the state $\left\vert h\right\rangle $, one can
known the quantum states of the atoms trapped in cavities A and B have been
directly exchanged.

Another immediate application is the heralded distribution of entanglement
between two atomic memories, which is shown in Fig. 6(b). A single-photon
pulse in the state $\left\vert v\right\rangle $ is divided into two paths by a
50/50 beam splitter (BS), one is reflected by a one-sided optical cavity
trapped an single atom with the quantum state $\left\vert r\right\rangle $.
The other is reflected by a mirror. The state evolution of this process can be
written as $\left\vert v\right\rangle \left\vert r\right\rangle \rightarrow
(\left\vert v\right\rangle \left\vert r\right\rangle +\left\vert
h\right\rangle \left\vert l\right\rangle )/\sqrt{2}$, which is an atom-photon
maximal entangled state. Then the state of the photon is stored into another
atomic memory by detection of the reflected photon. After the successful
generation of entanglement between two atomic memories within the attenuation
length, one wants to extend the quantum communication distance. This is done
through entanglement swapping \cite{Briegel,Duan,Razavi,Simon}. Suppose that
we start with two pairs of entangled memories. First, one of each entangled
pair is mapped into single photon with retrieval operation. Second, two
photons from two pairs of entangled ensembles are performed Bell-state
measurement with two photon interference \cite{Huang}. After an entangled
state has been established between two distant memories, we can use it in
entanglement-based communication protocols, such as quantum teleportation,
cryptography, and Bell inequality detection with linear optics elements
\cite{Duan}, since the quantum states between atomic memories and photons can
be transferred in a reversible manner \cite{Kimble}.

Next we briefly address the experiment feasibility of the proposed schemes. We
consider a $^{87}$Rb atom trapped in a one-sided Fabry--Perot cavity
\cite{Sauer}. The states $\left\vert l\right\rangle $ and $\left\vert
r\right\rangle $ correspond to $\left\vert F=1,m=-1\right\rangle $ and
$\left\vert F=1,m=1\right\rangle $ of $5S_{1/2}$ ground levels, respectively,
while $\left\vert e\right\rangle $ corresponds to $\left\vert
F=1,m=0\right\rangle $ of $5P_{3/2}$ excited level. The relevant cavity QED
parameters for this system are assumed as $(g_{0},\kappa,\gamma_{e}%
)/2\pi=(27,4.8,6)MHz$ \cite{Sauer}, which fit well the condition $4g_{0}%
^{2}/\kappa\gamma_{e}\sim102\gg$ $1$. Suppose that the average photon
absorption rate in a fiber is $1-e^{-L/L_{att}}$, where $L$ is the length of
the fiber and the channel attenuation length $L_{att}\sim22km$ \cite{Simon}.
If the single-photon detection efficiency is $\eta_{d}\sim0.95$ \cite{Simon},
the success probability for heralded quantum state or generation of
entanglement between two atomic memories with the distance $L\sim10km$ is
$P_{1}\sim P^{2}\eta_{d}e^{-L/L_{att}}\sim0.55$. For quantum communication
over long distance $L_{m}=2^{m}L$, one needs mth (m = 1, 2,...,N) entanglement
connection. The total probability to create entanglement across two
communication nodes at a distance of 1280 km is about $P_{1}%
{\displaystyle\prod\nolimits_{n=1}^{6}}
P_{n}\sim P_{1}%
{\displaystyle\prod\nolimits_{n=1}^{6}}
(\frac{1}{2}P^{2}\eta_{d}^{2}e^{-L/L_{att}})^{6}\sim0.0002$.

\emph{Conclusion}.---In summary, we have analyzed the heralded quantum memory
for single-photon polarization state with a single atom trapped in an optical
cavity \cite{Sauer}, and demonstrated its applications in quantum
communication network \cite{Kimble}. In our scheme, storage and retrieval of
photon polarization states have a high probability and fidelity. Heralded
storage, entanglement distribution, and quantum communication are achieved by
detecting the reflected photon. Thus in a realistic application operation
errors due to all sources of photon loss, including atomic spontaneous
emission, cavity mirror absorption and scattering, and photon collection, are
always signaled by the absence of a photon count. As a result, the photon loss
only decreases the success probability but has no contribution to the gate
infidelity if the operation succeeds (i.e., if a photon count is registered).
The ability to detect whether the operation has succeeded or not, is crucial
for practical application.

\textbf{Acknowledgments:} This work was funded by National Natural Science
Foundation of China (Grant No. 10574022 and Grant No. 60878059), the Natural
Science Foundation of Fujian Province of China (Grant No. 2007J0002), the
Foundation for Universities in Fujian Province (Grant No. 2007F5041), and
\textquotedblleft Hundreds of Talents \textquotedblright\ program of the
Chinese Academy of Sciences.

\end{document}